\documentclass[%
preprint,
prl,
 ]{revtex4-1}

\usepackage{graphicx}

\begin{document}
\title{An alloy calculation of pure antiferromagnetic NiO}

\thanks{Thanks to prof. Lucy Assali for very helpful discussions and to
CNPq, a Brazilian research supporting agency}%

\author{Luiz G . Ferreira}
\email{guima00@gmail.com}
\affiliation{Instituto de F\'isica, Universidade de S\~ao Paulo, Brazil}
\begin{abstract}
We use the many techniques of alloy theory to study antiferromagnetic $NiO$, considered as
an alloy of spin-up and spin-down $Ni$ atoms. The questions are:
the true antiferromagnetic ground state and the possibility of obtaining ferrimagnetic configurations.
Further we use the GGA/LDA-1/2 technique to investigate the electronic excitation spectrum.
We found two valence bands and band gaps, of $\simeq 4.0 eV$ consistent with 
bremsstrahlung-isochromat-spectroscopy (BIS) result, and $\simeq1.2 eV$ consistent with the known
$10Dq$ value for the $Ni^{++}$ ion, and with the inelastic X-ray and energy-loss experiments.
The features of a Mott insulator are presented without recurring to an electron-pair
correlation.
\end{abstract}
\maketitle

\section{1. Introduction}
We applied the powerful techniques \cite{Fontaine,Sanchez} of alloy calculations to the
 antiferromagnetic Ni(II)O in the rock-salt structure.
We aim at verifying that the CuPt ($L1_1$) configuration of spins is truly the ground state, and
 verifying that no
ferrimagnetic arrangement is stable. The magnetic arrangement is described as an Ising alloy (ordered or
 not) of spin up and spin down $Ni$ atoms. First-principles calculation are used to determine the
  energies of prototypical configurations and cluster expansions (CE) are generated from these
   configurations. The 
cluster expansions (CE) allow predictions for the magnetic ground state.
\par Another interesting point related to our calculations (all made using the WIEN2k  LAPW code \cite{Wien}) is
 the spectrum of one-electron
excitations. So far, the official answer is that $NiO$ is a semiconductor with a large band gap
 ($\simeq 4.0 eV$) which is calculated with a LDA+U, GGA+U, or GW technique. There are many papers pointing to this result
\cite{gap-theo,Peter,Das,Eder} agreeing very well with the experimental band gap \cite{gap-exper}.  On the other
 hand, it is
very well established the existence of much smaller gaps, meaning that there are other valence
 and/or 
conduction band extremes \cite{Merlin,Fromme,Huotari,Muller}. So we are also willing to investigate this
 possibility.
\section{2. Magnetic Configurations}
A first-principles calculation of antiferromagnetism and ferrimagnetism is not always simple. The
 procedure
coded in WIEN2k is not always useful for our purposes. We present
a new procedure based on alloying theory. We consider a ferri or antiferromagnet as an alloy of
 spin-up and
 spin-down atoms in a lattice. The calculation is spin-polarized and most were made in two steps.
  In the first
 step we add an attractive potential of perturbation to the atoms of spin up and a repulsive
  potential to the
  atoms of spin down, for the solution of the Schroedinger equation of spin-up electrons. For the
   Schroedinger
   equation of spin-down electrons we add a repulsive potential for the atoms of spin up and
    attractive for 
    the spin down atoms. The calculation is made self-consistent and usually attains the
     magnetization that we
 want. In the second step we remove the added potentials and run the self-consistent cycles again.
  The result
  is an unperturbed magnetic structure, either antiferromagnetic, ferrimagnetic or ferromagnetic
   usually
   according to the planned distribution of magnetic atoms. It might happen that the magnetic
    ordering of the
    final state is not the one that was planned, but this was an exception never verified in our
     $NiO$  calculations. 
\par
\begin{figure}[h]
 \includegraphics[scale=0.5]{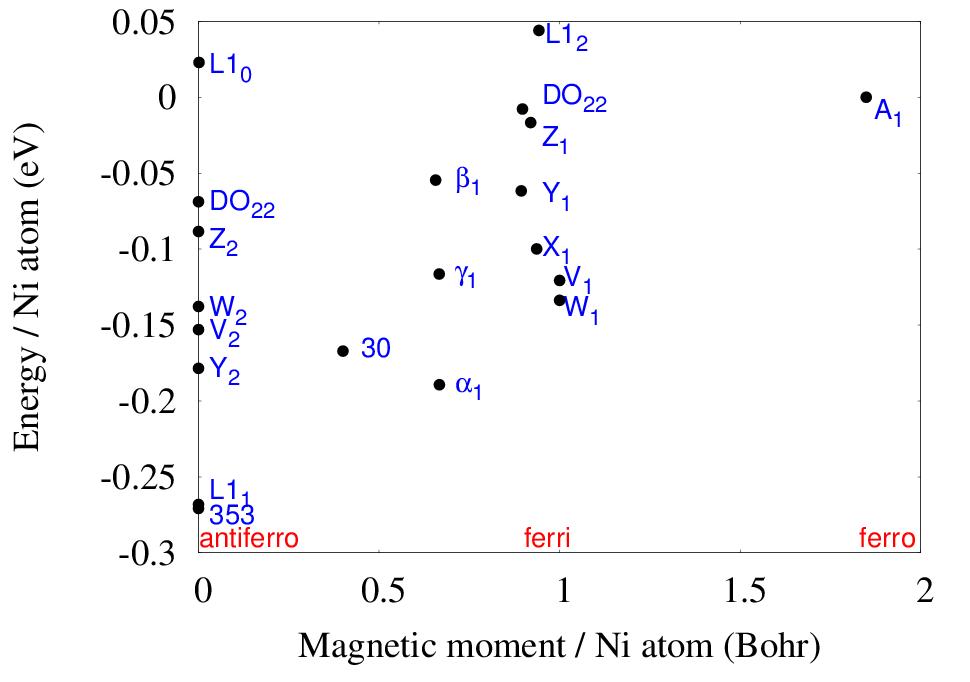}
\caption{(Color Online)Energy and magnetic moment of PBE/LAPW calculated configurations. Configurations with
zero magnetic moment are exactly antiferromagnetic. The configuration with moment $\simeq 2$ is
 ferromagnetic.
In between one has many possible ferrimagnetic states. The energy zero is the ferromagnetic arrangement 
$A_1$.} 
\label{EvsM}
 \end{figure}

 Fig.~\ref{EvsM} illustrates the magnetic moment and energy calculated for prototypical
 configurations.
Except for the configurations $30$ and $353$ they were all used to find the cluster expansions
 (CE). These
configurations and their symbols follow the references 
\cite{US-BR,Laks,gang,gang2}. These energies and momenta were calculated with LAPW using PBE \cite{PBE}
exchange-correlation energy. 
The prototypical configurations have at most four $Ni$ atoms per cell. Aside from these
 configurations we
 calculated configuration $30$, which has 5 $Ni$ atoms in the cell, and configuration $353$ with 8
  Nickel atoms. Configuration $30$ has no importance but was calculated nonetheless
  \cite{footnote}. Configuration
   $353$ was found
 to be the ground state of rock-salt antiferromagnetic $NiO$, degenerate with
 configuration $L1_1$. 
\subsection{2.a. Cluster expansions}
For alloys one uses the well known cluster expansion (CE) \cite{Sanchez}
\begin{equation}
E(\sigma)=\sum_{j,k}J_j\Pi_{j,k}(\sigma)=\sum_jJ_j\bar{\Pi}_j(\sigma)
\label{CE}
\end{equation}
where $E$ is the energy (per $Ni$ atom) of a configuration $\sigma$ of atoms $A$ and $B$ or spins $up$
 and $down$
in a lattice. $\Pi_{j,k}$ is a product of Ising spins $S=1 \ or\ -1$ at the vertices of a polyhedron $j$ drawn
 in the
 lattice. $\Pi$ depends on the configuration. There are tables and codes for the calculation of
  $\Pi_{j,k}$ \cite{Ferreira,Ceder}. $k$
  numbers identical polyhedron displaced by translation or rotation symmetry operations, and
  $\bar{\Pi}_j$ is the average of $\Pi_{j,k}$ in the set of identical polyhedra
  \cite{Laks}. $J_j$ are the $Ni$-$Ni$ interaction parameters.  In the case of $NiO$ we have 18 configurations in
   our data set, thus we can find at most 18
interaction parameters $J_j$, assuming all the other interactions are negligible. 
\par The whole procedure to find the CE is the following. Assume we have first-principles
 calculated more 
configurations than the number of $J_j^{'}$s we plan to use in our CE. Then
we find the  $J_j^{'}$s by least square error fit . In many instances this procedure will lead to a very
 wrong CE, and we must have a recipe to choose the size of the CE and which interactions $J_j$ to
  use.
The recipe was formulated in the reference \cite{Ceder} and frequently leads to short CE's
 \cite{Guima et al}.
Reference \cite{Ceder} uses
a figure of merit that corresponds to the predictive power of the
set of interactions. The idea is the following. Let $\sigma$ be a
configuration of the set, let $e(\sigma)$ be its first-principles
total energy per $Ni$ atom,
  let $j$ be an interaction (figure) of the set, and  $J_j$ its value, and let
$E(\sigma)$ calculated according to Eq.~\ref{CE} 
be the cluster expansion approximation to the true value $e(\sigma)$.
\par If the set of interactions and the set of configurations are given,
the interaction values $J_j$ should be chosen so to minimize the rms error
\begin{equation} rms^2=\frac{1}{N}\sum_{\sigma=1}^N[e(\sigma)-E(\sigma)]^2
=min.\label{min}\end{equation}
This minimization brings no information on the predictive power of
the set of interactions. To know its predictive power we consider
the set of configurations with one of them excluded, say configuration
$\omega$. With this exclusion we recalculate the values $J_j$, again
using Eq.~\ref{min}, and obtain the approximation  $\hat{E}(\omega)$
to the first-principle calculated value corresponding to the excluded
configuration.
Following reference \cite{Ceder} we define the `cross-validation' (CV)
figure of merit as
\begin{equation}\label{CV}
CV^2=\frac{1}{N}\sum_{\omega=1}^N[e(\omega)-\hat{E}(\omega)]^2
\end{equation}
in other words, we sum squared errors for each configuration when
it is excluded from the set. As a practical way to calculate $CV$ one proves
the relation
\begin{widetext}
\begin{equation}\label{CV2}
e(\sigma)-\hat{E}(\sigma)=[e(\sigma)-E(\sigma)]\left[1-\sum_{j,m}\Pi_j(\sigma)
Q(j,m)^{-1}\Pi_m(\sigma)\right]^{-1}
\end{equation}
\end{widetext}
where $Q$ is the matrix
\[Q(j,m)=Q(m,j)=\sum_{\sigma}\Pi_m(\sigma)\Pi_j(\sigma).\]
Eq.~\ref{CV2} shows that $CV$ is always greater than $rms$ error.
\par In the case of $NiO$ we started from the first 13 nearest neighbour pair interaction 
and the four-body nearest neighbour interaction (a regular tetrahedron of lattice sites). 
The cross-validation $CV$ was decreased when we reduced the number of interactions. We ended
with one CE with the first two nearest-neighbour pair interactions, named $J_2$ and $K_2$, and 
the tetrahedron interaction $J_4$. The labels and definitions of these interactions follow
references \cite{US-BR,Laks,gang,gang2}.
For this CE, the cross-validation was $CV=0.01384~eV$ and the root-mean square error
was $rms=0.01244~eV$. It is amazing that longer-range pair interactions only damage the
CE. For comparison, this CE predicted an energy of $-0.2705~eV$ for the data-set configuration
$L1_1$ while the LAPW result is $-0.2662~eV$. The zero of energy being used is the
ferromagnetic $A_1$ configuration in Fig.~\ref{EvsM}.
\subsection{2.b. The ground state}
\par Using the CE and scanning our file of configurations, which has all configurations
up to 8 $Ni$ atoms per cell, we found a configuration with number $353$ which, together with
configuration $L1_1$ is the ground state of rock-salt $NiO$. This same result was obtained with a 
CE without the four body interaction but with 3
 pair interactions instead of 2. This latter CE had slightly larger $CV$. The CE predicts 
  $E($353$)=-0.2873~eV$ and the all-electron LAPW   code gives
$E($353$)=-0.2708~eV$. The true ground state, $353$ or $L1_1$, depends on 
the parameters of the calculation, such as the size of the wave-function, charge density and
potential expansions,
exchange-correlation approximation and could not be determined. In all cases the CE results
are consistent with first-principles. For instance, using the exchange of Ref. \cite{PBEsol} 
the difference between the energies per $Ni$ for $L1_1$ and $353$ is only $0.0006~eV/Ni$.
This degeneracy is not related to symmetry
which is very different for the two configurations. $353$ has space group $227\  Fd-3m$,
while $L1_1$ is rhombohedral ($166\  R-3m$). Each $Ni$ atom of configuration $353$ has 6 first-neighbours
with the same spin and 6 with opposite spin, as in the configuration $L1_1$, though with
a different space distribution. The other antiferromagnet of Fig.~\ref{EvsM} do not have this 6/6
 distribution of neighbours.
\par Configuration $L1_1$ is an alternation of spin-up and spin-down planes along the cubic direction
$(111)$. Its energy may be lowered by a shear strain deformation along this direction. The gain
in energy is in the order of a fraction of $meV$, thus unable to decide on the ground state. 
Configuration $353$ is highly symmetrical, with space group Fd-3m. In units of a
simple cubic lattice parameter the atomic positions are:
\begin{itemize}
\item Spin up: (0.5,0,0); (0.25,0.25,0.5); (0.75,0.5,0.75); (0,0.75,0.25)
\item Spin down: (0.25,0,0.25); (0.5,0.25,0.75); (0,0.5,0.5); (0.75,0.75,0)
\end{itemize}
One readily sees that, with respect to a vector (210), the spin up atoms occupy planes
positioned at z=-0.25, 0., 0.75, 1., 1.75, 2., ...
and spin down are at z=0.25, 0.5, 1.25, 1.5, 2.25, ...
The 12 $Ni$ neighbours of each $Ni$ atom in configuration $353$ are in two octahedra, one made of spins up, 
the other with spins down. 
 \begin{figure}[h]
 \includegraphics[scale=0.5]{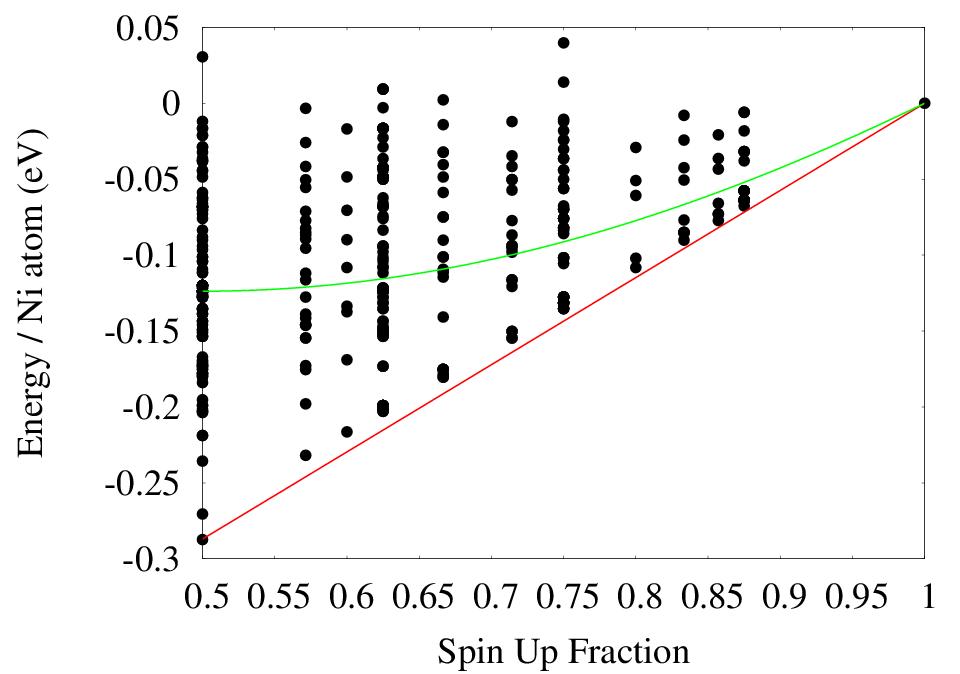}
\caption{(Color Online)Energy for each of the 365 configurations in the file containing all with
up to 8 $Ni$ atoms per cell against the fraction of spin up $Ni$ atoms. The straight line joining
the antiferromagnetic ground state at $x=0.5$ to the ferromagnetic $x=1$ means
the locus of the two-phase mixtures of anti- and ferromagnet. The curved line
is the paramagnetic phase with spins randomly oriented. Clearly all ferrimagnetic
solutions decay into the two-phase mixture showing the impossibility of ferrimagnetic $NiO$}
\label{ground}
 \end{figure}
 \par Fig.~\ref{ground} hints at the impossibility of $NiO$ presenting a ferrimagnetic phase.
 As explained in the Fig., such phase would decay into a two-phase mixture. The argument
 is based on our file containing 365 configurations. Ordered configurations with more than
 8 Nickel atoms per cell would be difficult to prepare, either in Nature or
  in a  Lab.
\subsection{2.c. Monte Carlo results}
\begin{figure}[h]
 \includegraphics[scale=0.5]{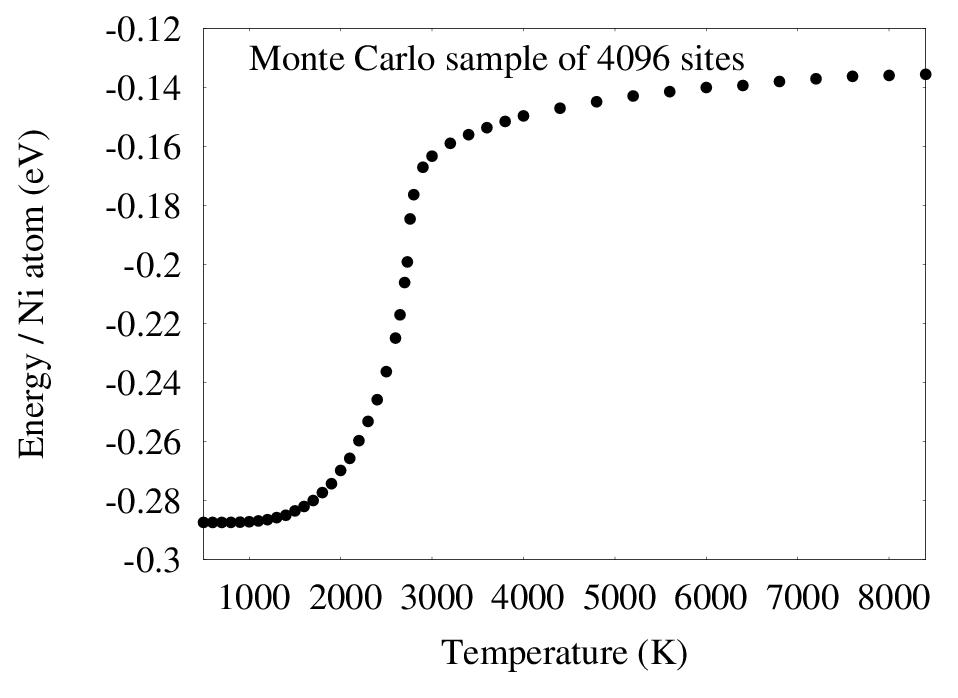}
\caption{Energy as function of $T$ for the Monte Carlo runs. The phase transition happens
at $T=2750\pm10 K$. The phase at low temperatures is certainly $353$, not $L1_1$, resulting
from the present choice of first-principles calculation parameters.  One
distinguishes the two phases by the value of $\bar{\Pi}(\sigma)$ corresponding to the
nearest-neighbour tetrahedron of sites. The calculated transition temperature is much too high,
even higher than the melting temperature.
Two reasons concur to that: 1 - We are using an Ising Hamiltonian, not Heisenberg For instance,
2D Ising models have ordered phases, Heisenberg models have not because the transverse spin
components allow paths of relaxation. 2 - We are assuming that sublattice magnetizations
are unique and do not depend on spin configuration of the lattice, that is, given the 
concentration $x$ the magnetization is known. Fig.~\ref{EvsM} shows the real situation where
the magnetization can fluctuate to some extent for configurations with the same $x$.}
\label{MonteCarlo}
 \end{figure}

Having the cluster expansion parameters, it is not difficult to run Monte Carlo calculations, 
following the Metropolis  algorithm \cite{Metropolis}. Fig.~\ref{MonteCarlo} shows the Energy
(Enthalpy) function of temperature. At low temperatures, the ground state is $353$,
not $L1_1$, for the present calculation, 
or a phase with the same $\bar{\Pi}$ for the nearest-neighbour pair, 
for the second-neighbour
pair, and for the tetrahedron first-neighbour 4-body interaction. Ising model leads to a much 
too high transition temperature, even higher than the melting point. In the Fig. 
 caption we present two reasons why the Ising model fails to determine the Thermodynamics.
\section{3. Electronic Excitations}
\par The literature on the electronic excitations of $NiO$ is very rich. It was accepted as a Mott
insulator and many theoretical methods were applied, to account for the strong correlations, and
 based on band calculations \cite{gap-theo,Peter,Das} or based
on cluster calculations \cite{Eder}. Without reviewing the many works
and techniques, we decided to give a chance to a very successful technique we developed
for the calculation of semiconductors: LDA/GGA-1/2 \cite{Guima et al,LDA-1/2,AIP}.
It is not unusual that  LDA/GGA-1/2 gives better results than GW or HSE \cite{bando}
We expect from our calculation: 1 - to produce an insulating $NiO$; 2 - a band gap of 
about 4.0~eV; 3 - smaller band gaps in the order of 1.0~eV. Expectations 2 and 3 are
incompatible, unless more than one valence  band is playing in the excitations.
\par Pure Kohn and Sham (KS) methods are good for the calculation of total energies, as we used
throughout the preceding section. When it comes to the calculation of excited states, the KS band
structures present very important errors. In the case of semiconductors, one misses band gaps
for the small gap semiconductors and, generally, the KS gaps are smaller than true gaps and
effective masses are lighter. These facts are universally recognized, and that is the reason
for the increased use of $U$ in LDA+U, GGA+U, etc. Despite of this fact one sees attempts
of performing single shots band calculations, simultaneously giving the total energy
and the excitation spectrum. A recent attempt is the work of Tran et al \cite{Tran}
that uses exact-exchange as a reference method of calculation. At this point it is
well to remind that half ionization methods beats true-exchange (Hartree-Fock)
by a very large margin \cite{AIP}.
\par
\begin{figure}[h]
\includegraphics[scale=0.5]{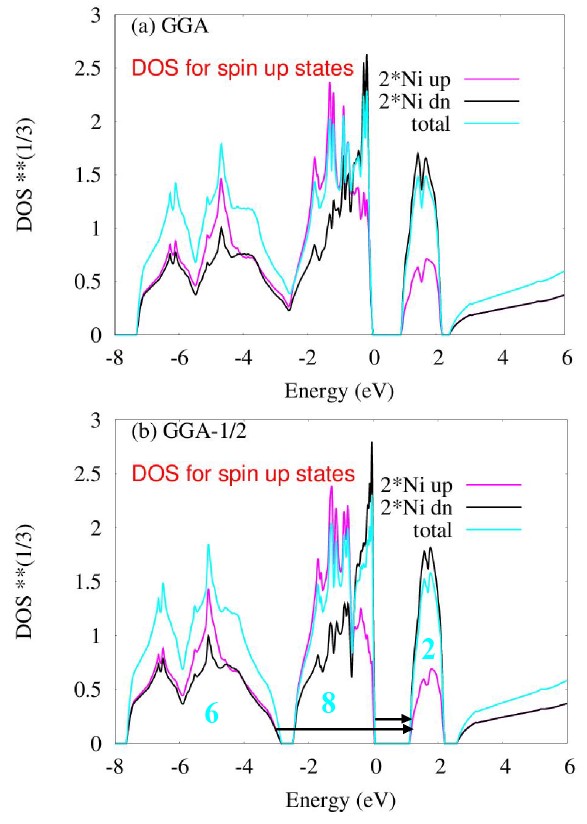}
\caption{(Color Online)Total DOS and partial DOS projected on the $Ni$ atoms.
The top panel (pure GGA(PBE)) equals the results of ref. \cite{Blaha2}. In the lower panel,
 the arrows indicate possible excitations. The numbers below the curves indicate the
 number of spin-up states in each pack. The first pack is mostly made of $O-p$ electrons. There
 are 6 such electrons with spin-up in this pack corresponding to the 2 $O$ atoms. 
 The second pack is made of 5 $Ni-d$ electrons of the
 atom with spin-up plus 3 electrons of the atom with spin down. The third pack is already
 a conduction band and it is made of two states in the spin-down atom. As always, 
 the cubic field splits the d-level into $3+2$ states, of which 3 states remain occupied and
 2 states are empty forming the first conduction band. One notices that the first conduction
 band is very narrow, meaning that it conducts poorly. }
 \label{DOS}
 \end{figure}

 Applying LDA/GGA-1/2 to $NiO$ is not straightforward. First we decided that the configuration 
to be investigated was the $L1_1$. 
We also made LDA/GGA-1/2 band calculations on the $353$ configuration but the results were wholly
similar to those of $L1_1$.
Secondly we must decide which atom, $Ni$ or $O$ should be 
half-ionized, and we chose the anion $O$, as is done for most LDA-1/2 calculations so far. 
Thirdly, since there are two $O$ in the cell, the "self-energy potential" was halved for each $O$,
as it is the usual practice, and multiplyed by $1/8$ in the case of 353. 
In the case of $NiO$, Fig.~\ref{DOS} shows the density of states for the pure GGA and
for the GGA-1/2. The pure GGA result coincides with that of ref. \cite{Blaha2} 
and does not account for
the $4.0 eV$ band gap \cite{gap-exper}.
On the other hand the GGA-1/2 maintains the $1.0 eV$ gap and opens a gap in the valence band.
 The method
has one free parameter chosen to maximize the band gap. The free parameter $CUT$ was chosen 
to maximize the band gap between the first valence band  and the 1st conduction band, that is
the long horizontal arrow in the Fig.~\ref{DOS}. The opened gap between
the 1st valence and the 2nd valence exists in the region of $1.6\leq CUT \leq 3.7~a.u.$. Opening a
gap is very common for LDA/GGA-1/2. In fact, most small gap semiconductors only present
the band gap in the LDA/GGA-1/2, because in the pure GGA or LDA they are metals. It is simple to
understand the two gaps: a) the gap of $4.0eV$ is the minority spin excitation
$Ni^{++}O^{--}\longrightarrow Ni^+O^-$; b) The gap of $\simeq1.0eV$ is the
minority spin excitation $t\longrightarrow e$.
\par Our results were the following. Band gap between 1st valence band and 1st conduction
band: 4.02~eV, in agreement with experimental BIS data. Band gap between 2nd valence band
and 1st conduction band: 1.18~eV, consistent with the $10Dq$ for $Ni^{++}$. 
Observe that the 1st conduction band is very narrow 
meaning that the excited states in this band are much localized in the spin-down $Ni$ atom. 
\par In what sense $NiO$ is a Mott insulator \cite{Mott}? First, aside
from being a semiconductor. it is a very poor
conductor, as one sees from the very narrow conduction band. Probably the conduction 
is mostly by holes in the first valence band. Second, the first band gap is
a $Ni\  3d \rightarrow Ni\ 3d$ transition which is optically forbidden. These features
come from the GGA-1/2 bands and do not require any assumption on the electron-pair
interaction.
 The band gap between the highest valence and the first
 conduction bands corresponds to $10Dq\simeq 1.0 eV$ of splitting $d$-states by a cubic field,
 compatible with the standard value for the $Ni^{++}$ ion in aqueous solution \cite{10Dq}.
 In studying these curves, one has to pay attention
 to the following facts: 1 - what is being plotted is the cubic root of the DOS, not
 the DOS itself; 2 - the partial DOS for the $Ni$ atoms is being doubled.
 \par It must be mentioned that this application of LDA/GGA-1/2 to $NiO$ is not 
 the first we made. In 2009 we presented results of another calculation \cite{old} where both
the  $Ni$ and the $O$ atoms were half-ionized. In that case there was no gap separating
the two parts of the valence band. We much prefer the present results because it is 
calculated with the most standard techniques within GGA-1/2. Further, the conduction band
at $4.0eV$ (or $1.0 eV$ counted from the top of the valence band) is d-like in the 
present version instead of s-like of the older version.
\section{4. Summary}
In this work we made an unusual study of $NiO$, considered as an alloy of spin-up
and spin-down $Ni$ atoms. We could not find a stable ferrimagnetic phase, which is
satisfying because no such phase was ever detected. But we were able to calculate an
antiferromagnetic phase $Ni_8O_8$ degenerate  with the $L1_1$ phase, for all 
practical purposes.
\par In the second part of this work we restudied the one-electron excitations by
means of the LDA/GGA-1/2 method. We used the most standard procedures within that 
method and found two band gaps, corresponding to a split of the valence band, 
and features of a Mott insulator. Hopefully, our 
results again match experiment.


\begin{thebibliography}{99}
\bibitem{Fontaine} D. de Fontaine, in {\it Solid State Physics}. edited by H. Ehrenreich,
F. Seitz, and D. Turnbull (Academic, New York, 1979), Vol. 34, p. 73.
\bibitem{Sanchez}J. M. Sanchez, F. Ducastelle, and D. Gratias, Physica {\bf 128A} 334 (1984).
\bibitem{Wien} P. Blaha, K. Schwarz, G. K. H. Madsen, D. Kvasnicka, J. Luitz,
"An Augmented PlaneWave + Local Orbitals Program for Calculating Crystal Properties",
Techn. Universit\"{a}t Wien, Getreidemarkt 9/156, A-1060Wien/Austria (2012).
\bibitem{gap-theo}. C. Toroker, D. K. Kanan, N. Alidoust, L. Y. Isseroff,
P. Liaob and E. A. Carter, Phys. Chem. Chem. Phys.,  {\bf 13}, 16644-16654 (2011);
M. C. Toroker and E. C. Carter, J. Mater. Chem. A {\bf 1} 2474-2484 (2013).
\bibitem{Peter} F. Tran and P. Blaha, Phys. Rev. Lett. {\bf 102}, 226401 (2009).
\bibitem{Das}Suvadip Das, John E. Coulter, and Efstratios Manousakis, Phys. Rev. B {\bf 91}, 115105 (2015).
\bibitem{Eder} R. Eder, Phys. Rev. B {\bf 78}, 115111 (2008).
\bibitem{gap-exper}G. A. Sawatzky and J. W. Allen, Phys. Rev. Lett. {\bf 53}, 2339 (1984).
\bibitem{Merlin} R. Merlin, Phys. Rev. Letters {\bf 54} 2727 (1985).
\bibitem{Fromme}B.Fromme, ``D-d excitations in transition metal oxides: a spin-polarized electron
 energy-loss spectroscopy (SPEELS) study'' (Springer-Verlag, Berlin-Heidelberg) 2001.
\bibitem{Huotari} S. Huotari, T. Pylkk\"{a}nen, G. Vank\'{o}, R. Verbeni, P. Glatzel, 
and G. Monaco, Phys. Rev. B, {\bf 78} 041102(R) (2008) .
\bibitem{Muller} F. M\"{u}ller and S. H\"{u}fner, Phys. Rev. B, {\bf 78} 085438 (2008).
\bibitem{US-BR} Z. W. Lu, S.-H. Wei. A. Zunger, S. Frota-Pessoa, L. G. Ferreira, Phys. Rev. B {\bf 44} 512 (1991-II).
\bibitem{Laks} D. B. Laks, L. G. Ferreira, S. Froyen, A. Zunger, Phys. Rev. N {\bf 46} 12587 (1992-I).
\bibitem{gang} V. Ozoli\c{n}\v{s}, C. Wolverton, and A. Zunger, Phys. Rev. B {\bf 57}, 6427 (1998).
\bibitem{gang2} L. G. Ferreira, V. Ozoli\c{n}\v{s}, and A. Zunger, Phys. Rev. B {\bf 60} 1687 (1999).
\bibitem{PBE}J. P. Perdew, S. Burke, and M. Ernzerhof, Phys. Rev. Let. {\bf 77} 3865 (1996).
\bibitem{footnote} Because of a defective cluster expansion that pointed to its importance.
\bibitem{Ferreira} L.G. Ferreira, S.-H. Wei e A. Zunger, Int. J. of Supercomputer Applications
 {\bf 5}, 34-56 (1991)
\bibitem{Ceder}A. van de Walle and G. Ceder, J. Phase Equilib. {\bf 23}, 348 (2002).
\bibitem{Guima et al} L. G. Ferreira, M. Marques, L. K. Teles, Phys. Rev. B {\bf 74}, 075324 (2006).
\bibitem{PBEsol}J. P. Perdew, S. Kurth, J. Zupan, and P. Blaha, Phys. Rev. Let. {\bf 82}, 2544 (1999).
\bibitem{Metropolis} N. Metropolis, A. W. Rosenbluth, M. N. Rosenbluth, A. H. Teller,
and E. Teller, J. Chem. Phys. {\bf 21}, 1087 (1953).
\bibitem{Mott}  N. F. Mott,  Proceedings of the Physical Society A {\bf 62},  416 (1949).
\bibitem{LDA-1/2} L. G. Ferreira. M. Marques, L. K. Teles, Phys. Rev B {\bf 78}, 125116 (2008).
\bibitem{AIP}  L. G. Ferreira. M. Marques, L. K. Teles, AIP ADVANCES 1, 032119 (2011).
\bibitem{bando} O. P. Silva Filho, M. Ribeiro, Jr., R. R. Pel\'{a}, L. K. Teles, L. G. Ferreira,
and M. Marques, J. App. Phys. {\bf 114} 033709 (2013).
\bibitem{Blaha2} Walid Hetaba, Peter Blaha, Fabien Tran, and Peter Schattschneider1, 
Phys. Rev. B {\bf 85}, 205108 (2012).
\bibitem{Tran}Fabien Tran, Peter Blaha, Markus Betzinger, Stefan Bl\"{u}gel,
Phys. Rev. B {\bf 91}, 165121 (2015).
\bibitem{10Dq}Hydrated ion in solution. Ni$^{2+}$ = 8600 cm$^{-1}$. D. S. McClure, "Solid State Phys. Advances 
in Research and Applications", 9, 399 (1959). Quoted
by A. Abragan and B. Bleaney, "Electron Paramagnetic Resonance of Transition Ions", OUP Oxford (2012),  p. 378
\bibitem{old} L. G. Ferreira, L. K. Teles, and M. Marques, arXiv:0910.4485v1 [cond-mat.mtrl-sci]
23 Oct 2009.
\end{thebibliography}
\end{document}